\documentclass[sigconf]{acmart}

\usepackage{booktabs} 
\usepackage{wrapfig} 

\setcopyright{none}

\acmConference[WikiWorkshop2018]{Wiki Workshop}{April 2018}{Lyon, France}
\acmYear{2018}
\copyrightyear{2018}

\begin{document}
\title{A Wikipedia-based approach to profiling activities on social media}
\titlenote{Produces the permission block, and copyright information}
\subtitle{Extended Abstract}

\author{Christian Torrero, Carlo Caprini and Daniele Miorandi}
\affiliation{%
  \institution{U-Hopper srl}
  \streetaddress{via R. da Sanseverino, 95}
  \city{Trento}
  \state{Italy}
  \postcode{38122}
}
\email{{name.surname}@u-hopper.com}








\renewcommand{\shortauthors}{C. Torrero et al.}

\begin{abstract}
Online user profiling is a very active research field, catalyzing great interest by both scientists and practitioners. In this paper, in particular, we look at approaches able to mine social media activities of users to create a rich user profile. We look at the case in which the profiling is meant to characterize the user's interests along a set of predefined dimensions (that we refer to as categories). A conventional way to do so is to use semantic analysis techniques to (i) extract relevant entities from the online conversations of users (ii) mapping said entities to the predefined categories of interest. While entity extraction is a well-understood topic, the mapping part lacks a reference standardized approach. In this paper we propose using graph navigation techniques on the Wikipedia tree to achieve such a mapping. A prototypical implementation is presented and some preliminary results are reported. 

\end{abstract}

%
%

\ccsdesc[500]{Theory of computation~Design and analysis of algorithms}
\ccsdesc[300]{Theory of computation~Theory and algorithms for application domains}
\ccsdesc[300]{Theory of computation~Semantics and reasoning}
\ccsdesc[300]{General and reference~Experimentation}
\ccsdesc[300]{Applied computing~Marketing}

\keywords{user profiling, social media, semantic analysis}

\maketitle

\section{Introduction}
\indent Over the past few years, web portal owners and mobile application developers have devoted increasing efforts
in building rich user profiles, which can be conveniently used in order to provide personalized contents and offers.\\
\indent In this scenario, \emph{social login} is a relatively new channel for accessing valuable insights into the user's
interests. Social login is the practice of accessing a web service without creating a brand new username/password pair but 
by signing in making use of an already existing social network account. To give a concrete example, when seeking 
new employees, employers more and more frequently allow candidates to submit an application by means of their LinkedIn account 
rather than by uploading a CV and/or filling in a form after the creation of a dedicated account on the employer's web portal.\\ 
\indent Besides being the \emph{de facto} standard tool for authentication, social login --- as quickly remarked before --- 
may also be used to recover detailed information about the user's attitudes and preferences by gaining access to his/her 
social activities\footnote{It is worth remarking that the use of social login gives place to consent-based profiling, in that 
the user provides explicit consent to access her personal information; if properly complemented by information on how the data 
shall be used, for which purposes and how it can be accessed/modified/deleted, this allows to comply with privacy regulations, 
including EU-issues GDPR}. In fact, their typical long-lasting temporal span enables \emph{profiling}, i.e.,
the detection of the user's core interests and, therefore, allows for product and service recommendations far more tailored 
than those stemming from other (usually) extemporary actions on the Internet, like flight ticket purchases and hotel reservations. 
In this light, it is important to notice that such a profiling potential associated to social login remains nowadays largely 
unused and enabling its exploitation is one of the main goals of the present work.\\
\indent Starting from social activities, the process of profiling basically consists of associating each user with a set
of sectors to which his/her attention is usually focused. For example, if a user often mentions in his/her activities on 
social networks movies he/she has watched (or sport events he/she has attended), it can be deduced that such a user likes 
cinema (or sport).\\
\indent Schematically, profiling can be structured into two steps: first, topics --- movies or sport events in the 
previous example --- have to be extracted from the activities (together with their frequency) and, second, they have to be 
traced back or ``projected'' onto some categories --- ``Cinema'' or ``Sport'' in the case mentioned before --- in order 
to yield a quantitatively accurate portrayal of the user's interests.\\
\indent For a given user, the first step can be carried out by relying on standard open source and proprietary technologies, 
resulting in a so-called \emph{topic map} made up of couples of the form \emph{(topic, n. of counts)}, detailing the topics 
extracted from the social activities and the number of times each one was encountered.\\
\indent The second step is usually the hardest one: assessing not only whether a topic can be associated to a category
or not, but also evaluating the strength of said connection, displays all the difficulties --- above all, the lack
of ground truth --- intrinsic to a measurement of semantic relatedness.\\
\indent In this paper we outline an algorithm for projecting topics onto categories (also referred to as ``sinks'' in what follows) 
that makes use of the Wikipedia tree, taking advantage of its large coverage and of the structured knowledge it conveys. Before 
introducing the algorithm and evaluating its performances in the next sections, a couple of remarks are in order to avoid any 
possible confusion about the main features of this work. First, it is worth stressing once more that the actual goal \emph{is 
not measuring relatedness} --- though a large part of this paper will be devoted to it --- but rather \emph{user profiling}, for 
which measurement of relatedness is a preliminary, albeit important, step. Second, the measurement of relatedness this
work focuses on differentiates from the typical one since the couples of topics at stake here do not feature a generic hyponymy/hypernymy 
relation\footnote{In linguistic, the terms \emph{hypernym} and \emph{hyponym} are associated to the extent of the semantic field of a 
term compared to that of another one. More precisely, a term is hypernym (hyponym) to a second one if its semantic field is broader 
(narrower). For example, ``vegetable" is a hypernym of ``carrot" while ``carrot" is a hyponym of ``vegetable". Finally, two terms are 
co-hyponyms if they are not hypernym/hyponym one to the other: an example is given by the terms ``carrot" and ``potato".} (including the frequent
case where the two topics are co-hyponyms); on the contrary, the couples of topics \emph{solely} targeted by our model contain a term that lies, 
in principle, several levels of hypernymy higher than the other. In other words, our algorithm is essentially tailored for hypernymy-asymmetric 
situations --- the more pronounced the asymmetry (and this is the case occurring when profiling a user), the better it should perform. This feature 
will also have non-negligible consequences when evaluating the algorithm performances.\\ 

\section{Methods and Algorithms}
\hspace*{0.3cm} In the rest of this paragraph it will be assumed that a list of $N_s$ sinks $(S_1,S_2,\ldots,S_{N_s})$ has been
preset\footnote{The complete list of sinks we employ throughout this study includes ``Arts'', ``Cinema'', ``Cuisine'', 
``Culture'', ``Economics'', ``Entertainment'', ``Fashion'', ``Geography'', ``History'', ``Literature'', ``Music'', ``Nature'', 
``Philosophy'', ``Politics'', ``Religion'', ``Science'', ``Sport'' and ``Technology and applied sciences''.} and that the topic 
map of a user $u$ is available and made up of $N(u)$ couples of the form \emph{(topic, n. of counts)}; moreover, the entries of 
the $\text{i}^{\text{th}}$ couple will be labelled as $tp_u(i)$ and $cnt_u(i)$.\\
\indent In our model, the percentage --- or \emph{score} --- $pc(u,S_m)$ of $u$'s social activities that can be associated to the 
$m^{th}$ sink is given by

\begin{equation}
\label{eq1}
pc(u,S_m) = \sum_{i=1}^{N(u)}cnt_u(i)\cdot w[tp_u(i),S_m]/\mathcal{N}_R(u)\ ,
\end{equation}
\vspace*{0.1cm}

\noindent where $w[tp_u(i),S_m]$ is a weight measuring the strength of the relatedness of topic $tp_u(i)$ to sink
$S_m$ and where the normalization term $\mathcal{N}_R(u)$ reads

\vspace*{0.2cm}
\begin{equation}
\label{eq2}
\mathcal{N}_R(u) =  \sum_{m=1}^{N_s}\sum_{i=1}^{N(u)}cnt_u(i)\cdot w[tp_u(i),S_m]\ .
\end{equation}
\vspace*{0.2cm}

\indent The rationale behind Eq.(\ref{eq1}) is rather straightforward: the contribution of a topic $tp_u(i)$ to the score of a 
sink $S_m$ is proportional to the number $cnt_u(i)$ of times $tp_u(i)$ is encountered (the larger $cnt_u(i)$, the most
the topic contributes to $S_m$) and to the strength of the connection between $tp_u(i)$ and $S_m$ (the stronger 
$w[tp_u(i),S_m]$, the larger the contribution).\\
\indent The importance of each category in $u$'s activities on social media will be ranked on the basis of the
corresponding $pc(u,S_m)$ --- the higher $pc(u,S_m)$, the more important the sink $S_m$ --- and such a ranking will
provide a quantitative map of user $u$'s interests, i.e. yield its \emph{profile}.\\ 
\indent As pointed out in the Introduction, a preliminary step to compute the percentages $pc(u,S_m)$ is the evaluation
of the weight $w[tp_u(i),S_m]$ for each topic and each sink and this, in turn, is basically a measurement of relatedness
between $tp_u(i)$ and $S_m$. In our model, such a measurement will be carried out by relying on the graph (somehow inappropriately also referred 
to as ``tree" in what follows) underlying the Wikipedia ontology.\\ 
\indent The first graph-based approaches to measuring semantic relatedness date back to the end
of the 80's~\cite{Rada89}: since then, several algorithms relying on graph theory have been
proposed, among others those explained in~\cite{Wu94,Hirst98,Leacock98,Jarmasz03}. After its introduction in 2001, Wikipedia has been the object 
of an extensive research activity, featuring not only studies focusing on relatedness measurement (see~\cite{Ponzetto06,
Ponzetto07,Witten08}), but also several ones dealing with varied subjects like, for example, the detection of detrimental 
information~\cite{Segall13}.\\
\indent Irrespective of whether the Wikipedia tree was exploited or not, a common feature of all the above-mentioned 
graph-based studies for relatedness measurement is their general purpose, in the sense that they aim at assessing
the relatedness of couples of terms whose hypernymy relation is not specified a priori. On the contrary, as explained early on,
the focus of the present, Wikipedia-based work is more specific hypernymy-wise, since we are mainly interested in determining the connections 
between a generic topic and a series of hypernyms (sinks) that lie (several) levels higher than it in the Wikipedia tree. In principle, such a more
peculiar situation should allow for the usage of topological features of the tree that are correspondingly less generic --- for instance, features that are 
explicitly asymmetric (hypernymy-wise) with respect to the concepts whose relatedness has to be measured --- and that, consequently, should better 
capture the specific hypernymy relation of each topic-sink couple at hand here.\\ 
\indent The remarks above do not mean that the ingredients used in the methods available in the literature are unuseful in the present study:
in fact, there are tools that, notwithstanding their general-purpose scope, do seem to still be very valuable in the specific case under inspection here. A prominent
example is given by \emph{the length of the shortest path} (LSP) between the concepts whose relatedness has to be measured. In fact, it still
sounds reasonable to assume that the longer the shortest path, the less related the concepts to each other.\\
\indent However, bearing in mind that the eventual goal is the profiling of a user, it is the strength of the relatedness between a topic and a sink 
\emph{compared to the other sinks} that actually matters: this might make the LSP in itself insufficient to tackle the kind of relatedness measurement under inspection 
and prepares the ground for the introduction of those hypernymy-asymmetric tools --- one of them, at least --- we mentioned before.

\begin{figure}
\centering
\includegraphics[width=\columnwidth]{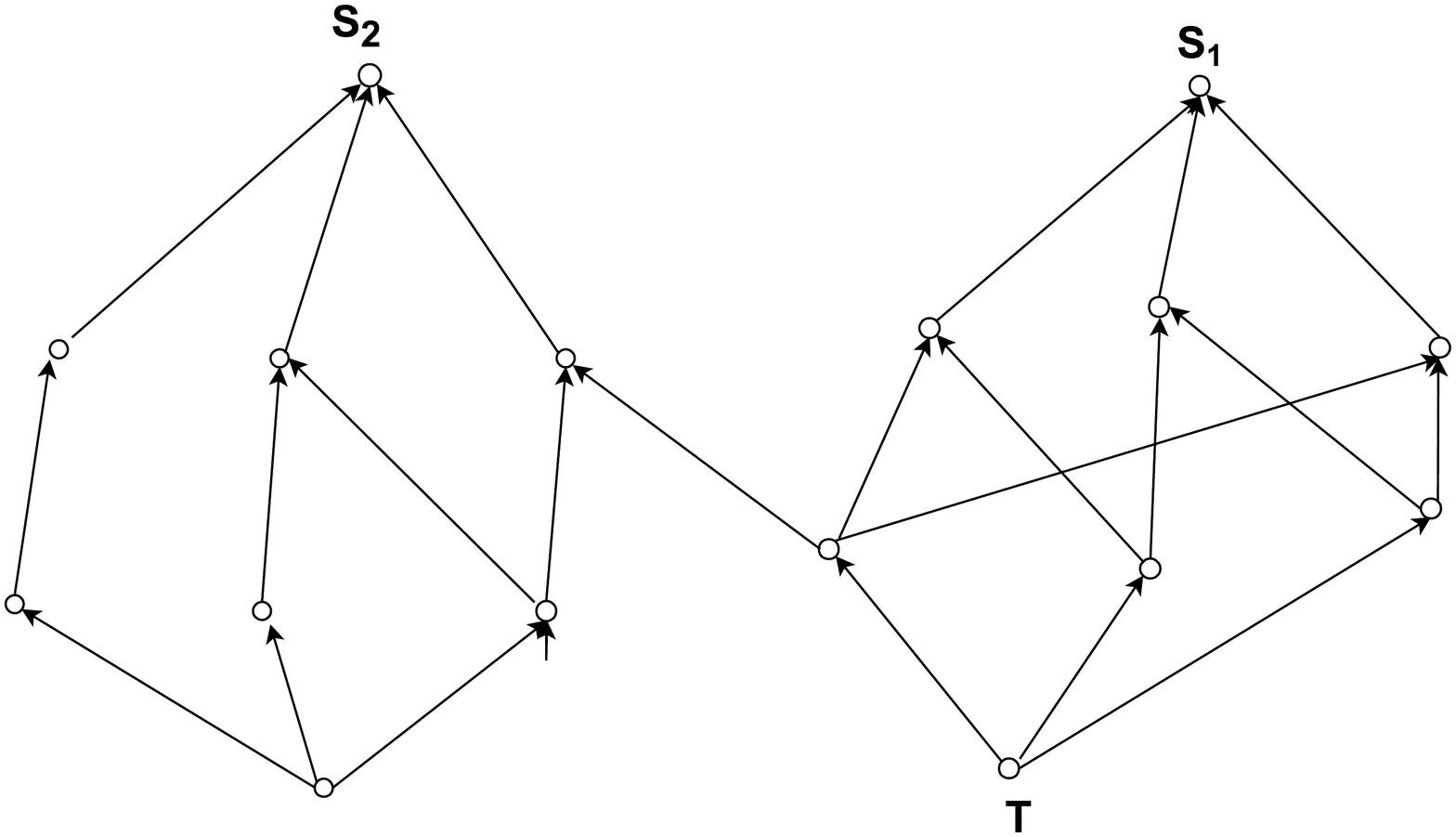}
\caption{Example of possible connections between a topic $T$ and two sinks $S_1$ and $S_2$. Arrows originating
from any node always point upwards, i.e. to the node hypernyms. The shortest path is 3 edges long for both sinks. In this and in all other
figures of this section, possible links to co-hyponyms are not shown at any hypernymy level to simplify the reading.}
\label{Fig1}
\end{figure}

\indent  As an example, consider Figure \ref{Fig1} where a portion of the Wikipedia tree is depicted, showing a topic $T$ that can be connected to two sinks 
$S_1$ and $S_2$. In such a figure, every link points to a node associated to a concept with a higher hypernymy with respect to the hypernymy of the term related 
to the node where the link originates. In other words, \emph{each link is oriented in a direction corresponding to a higher level of hypernymy}\footnote{Obviously, 
there might exist edges connecting co-hyponyms but they are not shown at any hypernymy level to simplify the reading of the graph. This applies also to all other 
figures in this section.}. In Figure \ref{Fig1}, the lengths $l_p(T,S_1)$ and $l_p(T,S_2)$ of the shortest path between $T$ and $S_1$ and $S_2$ respectively is 3 for 
both sinks. Consequently, no matter how $l_p(T,S_1)$ and $l_p(T,S_2)$ are combined, we would always argue that $T$ is related in the same way to both $S_1$ and 
$S_2$, as long as only $l_p(T,S_1)$ and $l_p(T,S_2)$ enter into play. However, the topology of the graph would suggest that the relatedness $w(T,S_1)$ between $T$ 
and $S_1$ should be stronger than that between $T$ and $S_2$. In fact, if one started from node $T$ and made random\footnote{It is assumed that, at each node along 
the way, all links that start from that node and go upwards are equally likely to be picked.} \emph{upward} moves (thus constantly increasing the hypernymy 
level), the probability of ending up in node $S_1$ would be higher than that of ending in node $S_2$ since there are more ``upward-pointing" paths 
connecting $T$ to $S_1$ than to $S_2$. By ``upward-pointing'' paths, we mean paths exclusively made up of links oriented in a direction corresponding to a 
higher level of hypernymy.\\
\indent Figure \ref{Fig1} suggests to take into account not only the length of the shortest path between a topic $T$ and a sink $S$ but also the overall number 
$n_p(T,S)$ of such ``upward-pointing" paths connecting $T$ to $S$: the larger $n_p(T,S)$,
the stronger the relatedness. Still in Figure \ref{Fig1}, one would have $n_p(T,S_1)=6$ and $n_p(T,S_2)=1$ and, therefore,
$T$ would turn out to be more strongly related to $S_1$ than to $S_2$, as expected. ``Upward-pointing" paths are an example of those topological
features, asymmetric in hypernymy, we were referring to early on as an instrument we could leverage to possibly improve the accuracy of
the specific kind of relatedness measurement targeted by this study.\\
\indent Considering again a given topic $T$ and recalling that there are $N_s$ sinks available altogether, the
formula we propose to measure the relatedness $w(T,S_m)$ of $T$ to the $m^{th}$ sink reads

\vspace*{0.15cm}
\begin{equation}
\label{eq3}
w(T,S_m) = \bigg(\frac{n_p(T,S_m)}{n_{tot}(T)}e^{-\alpha\big[l_p(T,S_m)-l_{MIN}(T)\big]}\bigg)/N_R(T)\ ,
\end{equation}
\vspace*{0.2cm}

\noindent where $\alpha>0$, $l_{MIN}(T) = min_i\{l_p(T,S_i)\ |\ \forall i\in[1,N_s]\}$,\newline
\noindent $n_{tot}(T) = \sum_{m=1}^{N_s} n_p(T,S_m)$ and where the normalization term $N_R(T)$ is given by

\vspace*{0.2cm}
\begin{equation}
\label{eq4}
N_R(T) = \sum_{m=1}^{N_s} \frac{n_p(T,S_m)}{n_{tot}(T)}e^{-\alpha\big[l_p(T,S_m)-l_{MIN}(T)\big]}\ .
\end{equation}
\vspace*{0.2cm}

\begin{figure}
\centering
\includegraphics[width=\columnwidth]{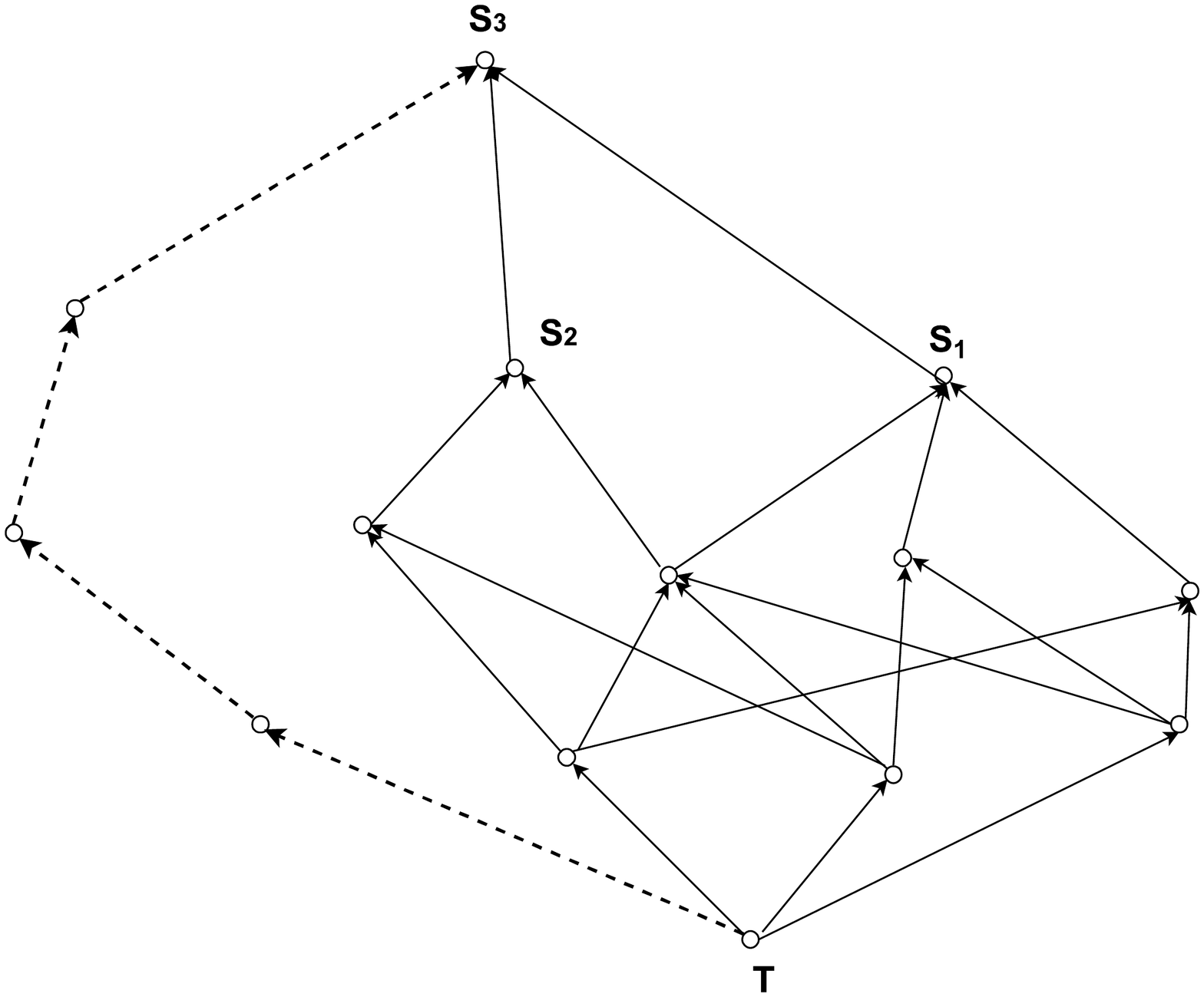}
\caption{Example of path dismissal. The only path connecting topic $T$ to sink $S_3$ is the dashed one since the remaining two
paths are discarded given that another sink is previously found along them.}
\label{Fig2}
\end{figure}

\indent According to Eq.(\ref{eq3}), any weight $w(T,S_m)$ results from the interplay between the number of ``upward-pointing" paths\footnote{From now, every time 
a path --- or a series of paths --- will be referred to, it will be tacitly understood that it is ``upward-pointing".} connecting topic $T$ to sink $S_m$ and the 
corresponding LSP. The latter topological feature actually enters not through its ``absolute" value but, rather, via its size relative to the minimum $l_{MIN}(T)$ of 
the lengths of the shortest paths between $T$ and any preset category. In this way, the strength of $w(T,S_m)$ somehow depends on the comparison between different 
sinks --- a key aspect when carrying out the profiling, as pointed out before.\\
\indent Some remarks are now due.\\
\indent First, in order to avoid to link a topic $T$ with a sink $S$ too far away, a maximal path length $l_{max}$
is introduced, i.e. paths between $T$ and any sink whose length is larger than $l_{max}$ are discarded in computing any weight $w$. It is true
that the relatedness between $T$ and a distant sink $S$ gets exponentially suppressed according to Eq.(\ref{eq1}), but $n_p(T,S)$
might be large in this case: thus, it might counterbalance --- partially, at least --- the exponential dump and result
in an unwanted (albeit small, perhaps) perturbation.\\ 
\indent As an example of this phenomenon taken from the Italian version of Wikipedia, let's consider the topic ``Thor", the god of Nordic 
mythology. If $l_{max}$ is set to 6 and the parameter $\alpha$ in Eq.(\ref{eq3}) is set to $3$\footnote{The reason for this choice --- 
and, more generally, the approach we followed in order to set the model parameters --- will 
be explained when assessing the performances of the algorithm in Section 4 of this paper.}, the only two categories (among those cited in footnote 3) whose weight is 
higher than 1\% are ``Religion" (scoring 88.2\%) and ``Politics" (11.8\%). While the former appears quite naturally (Thor is a divinity and, 
as such, he can be obviously associated to ``Religion"), the latter has a somehow less intuitive connection (``Politics" actually shows up since 
Thor is a god of war and war can be deemed as a political activity). With such a setup, i.e., $l_{max}=6$ and $\alpha=3$, 
$l_p(``Thor",``Religion") = l_p(``Thor",``Politics")=4$ while $n_p(``Thor",``Politics")=2$ and $n_p(``Thor",``Religion")=15$. If $l_{max}$ 
is increased to 10 while keeping $\alpha$ fixed, weights change since ``Religion" and ``Politics" now score 69.2\% and 30.2\% respectively. 
The reason is that, while $l_p(``Thor",``Religion")$ and $l_p(``Thor",``Politics")$ keep on being equal to 4, now $n_p(``Thor",``Politics")=7$ 
while $n_p(``Thor",``Religion")=16$. In other words, $n_p(``Thor",``Religion")$ essentially has not changed while $n_p(``Thor",``Politics")$ 
has become more than three times larger than it was before, thus resulting in an increase in importance of the spurious --- to some extent --- 
category. Consequently, in a case like this, setting $l_{max}$ equal to a short rather than to a large value does seem to yield more reasonable 
results.\\
\indent Second, in case there are no paths with length shorter than $l_{max}$ connecting a topic to any of the sinks, $l_{max}$ is increased 
by one unit \emph{temporarily and just for that topic} as long as at least a sink is encountered\footnote{An upper threshold $l_{th}$ to such 
a procedure is introduced to avoid paths unreasonably long.}: this avoids the lack of mapping for some topics and the resulting loss
of information about the user's interests.\\
\indent Third, once that a sink $S$ is encountered along a path, no further move upwards is considered along it, even though the
length of such a path leading to $S$ is less than $l_{max}$. For example, assuming $l_{max}=4$, in Figure \ref{Fig2}, the only path
connecting $T$ to sink $S_3$ is the one whose edges are dashed, since the remaining two links incident to $S_3$ are dismissed given
that they belong to paths --- starting from $T$ --- along which other sinks have already been found. The main reason for this choice is given by the fact that
sinks might not necessarily be at the same height along the tree, i.e. some sinks might belong to the hypernyms
of another sink. As an example, consider the sinks ``Music'' and ``Arts'': though the latter is a hypernym of
the former\footnote{With reference to Figure \ref{Fig2}, ``Music'' and ``Arts'' could be associated to sinks $S_2$ and $S_3$ respectively.},
the span of ``Music'' in the real world (that is, the space devoted to it on the media, the attention on the part of the
audience, the amount of resources invested in it, etc.) compared to that of, say, ``Ceramics'' is so much wider that we might very
well want to consider ``Music'' a sink on its own while merging ``Ceramics'' into the more generic ``Arts'' sink. In such a situation
any path connecting a topic $T$ to ``Music'' whose length is shorter than $l_{max}$ could easily be prolonged to reach ``Arts'',
therefore establishing a connection between the latter and $T$. Such a connection would not obviously be wrong but,
by increasing $w(T,\text{``Arts''})$, it will automatically weaken $w(T,\text{``Music''})$, which constitute an undesired effect given that
``Music'' is assumed to be a sink on its own.\\

\begin{figure}[htb]
\centering
\includegraphics[width=\columnwidth,height=8.0cm]{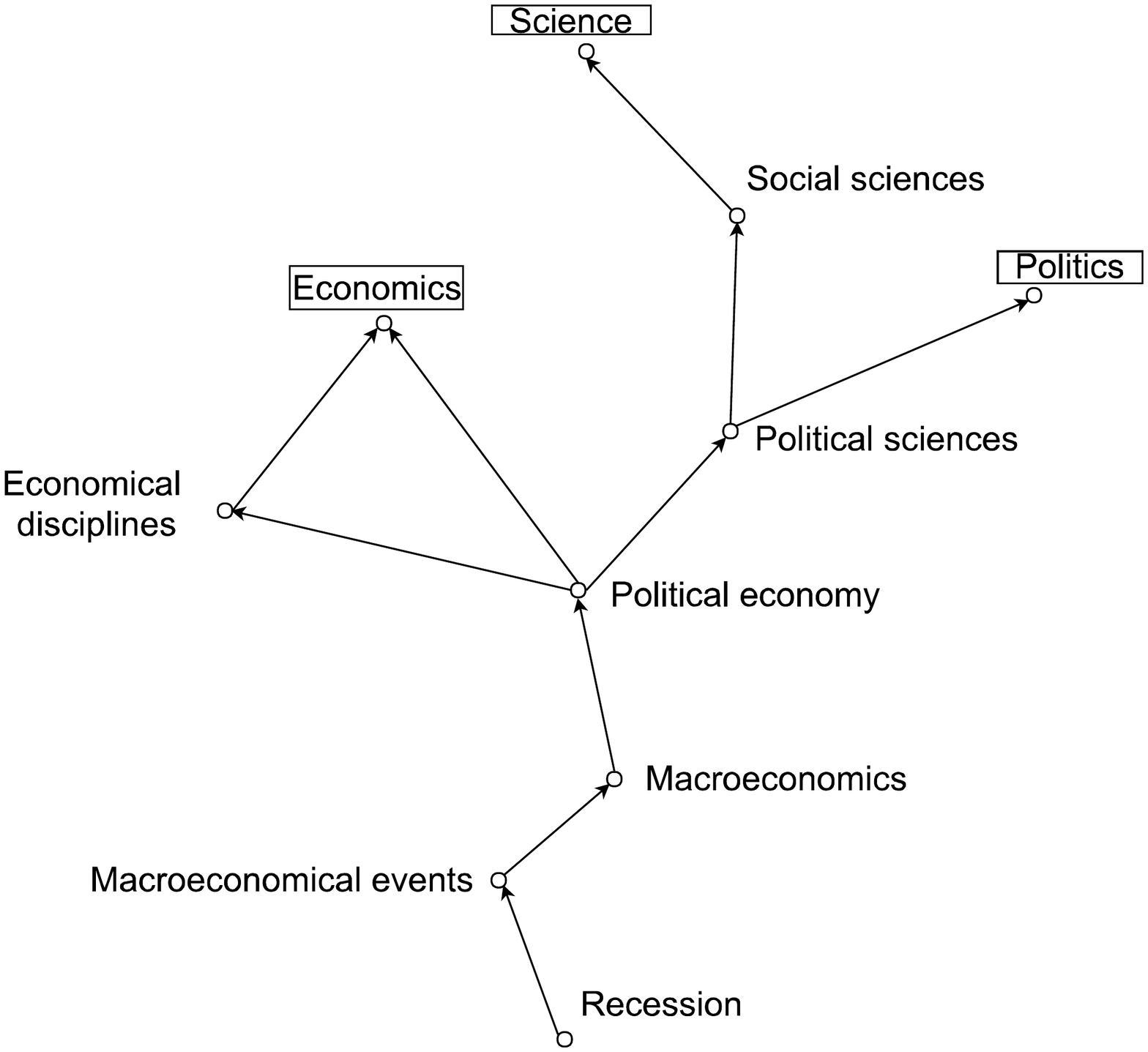}
\caption{Example of a computation of relatedness taken from the Italian version of Wikipedia. Starting from the node associated to the
concept of ``Recession'' (at the bottom), all edges belonging to paths leading to the sinks ``Economics'', ``Politics'' and ``Sciences'' are shown.
Other 15 sinks --- spanning a wide variety of fields and listed in footnote 3 --- were also considered but no paths leading to them and long 6 edges at most were 
found.}
\label{Fig3}
\end{figure}
\indent In summary, the algorithm for profiling a user $u$ starting from his/her topic map goes as follows:
\vspace*{0.2cm}
\begin{enumerate}
\item a list of sinks $(S_1,S_2,\ldots,S_{N_s})$ is built, a value for parameters $\alpha$ in Eq.(\ref{eq1}), $l_{max}$ and $l_{th}$ is chosen and
a cycle over the topics in $u$'s topic map is introduced;
\item for each topic $T$ in the cycle, paths between $T$ and the sinks with length not exceeding $l_{max}$ are built, bearing in mind
that any further move upwards along a path has to be dismissed as soon as a sink is encountered, irrespective of the current length of
the path;
\item in case no paths as those described in step 2 are found, $l_{max}$ is temporarily increased by one unit as long as either any valid
path connecting $T$ to at least one of the sinks is found or a given threshold $l_{th}$ is reached; afterwards, $l_{max}$ is brought back to its initial value;
\item the relatedness of $T$ to each sink is computed by means of Eq.(\ref{eq3});
\item finally, sinks are ranked on the basis of the scores defined in Eq.(\ref{eq1}). 
\end{enumerate}
\vspace*{0.2cm}
\indent As an illustrative example, let us consider the categorization of the concept ``Recession'' based on the Italian version of Wikipedia. The resulting 
graph is shown in Figure \ref{Fig3}, where, setting again $l_{max}=6$ and $\alpha=3$, we represent all the edges belonging to paths leading from the concept 
of ``Recession'' (``R'') at the bottom to the sinks ``Economics'' (``E''), ``Politics'' (``P'') and ``Sciences'' (``S''). No other sinks in the list detailed 
in footnote 3 can be reached with $l_{max}=6$. Making use of the notations introduced earlier, with reference to Figure \ref{Fig3} one has
$n_p(\text{``R''},\text{``E''})=2$, $n_p(\text{``R''},\text{``P''})=1$, $n_p(\text{``R''},\text{``S''})=1$, $l_p(\text{``R''},\text{``E''})=4$,
$l_p(\text{``R''},\text{``P''})=5$, $l_p(\text{``R''},\text{``S''})=6$ and, thus, $n_{tot}(\text{``R''})=4$ and $l_{tot}(\text{``R''})=4$. Consequently, the 
normalization factor $N_R(\text{``R''})$ and the relatedness $w$ of ``Recession'' to the three sinks are given by

\begin{eqnarray}
N_R(\text{``R''}) &=& \frac{1}{2} + \frac{1}{4}e^{-3} + \frac{1}{4}e^{-6} \approx 0.51\ , \nonumber\\
w(\text{``R''},\text{``E''}) &=& \frac{1}{2}/N_R(\text{``R''}) \approx 0.98\ , \nonumber\\
w(\text{``R''},\text{``P''}) &=& \frac{1}{4}e^{-3}/N_R(\text{``R''}) \approx 0.02\ , \nonumber\\
w(\text{``R''},\text{``S''}) &=& \frac{1}{4}e^{-6}/N_R(\text{``R''}) \approx 0\ .
\end{eqnarray}
\vspace*{-0.1cm}

\section{Implementation}
\hspace*{0.3cm} We developed a prototypical implementation of the algorithm devised in the previous section, in the form of a Python library using the functionality 
provided by our Tapoi\footnote{\url{http://www.tapoi.me/}} platform for the collection of online user activities and the generation and aggregation of the topic maps. 
Tapoi makes use of the commercial Dandelion APIs\footnote{\url{https://dandelion.eu/}} for entity extraction.\\
\indent To speed up the computation, a dump of the Italian version of the Wikipedia tree has been downloaded\footnote{We used the \texttt{categorylinks} (Wiki category 
membership link records) and \texttt{page} (base per-page data: id, title, old restrictions, etc.) taken from \url{https://dumps.wikimedia.org/itwiki/20180120/}} and 
stored in a standard relational database management system, with the Python code interfacing directly with said database, thereby bypassing Wikipedia APIs\footnote{\url{https://www.mediawiki.org/wiki/API:Main_page}}.\\
\indent Though at the moment the prototype makes use of the tree underpinning of the Italian version of Wikipedia, the analysis can be easily extended to other 
languages. In fact, Dandelion APIs automatically detect the language(s) used in a user's social media activities and set topic labels accordingly within the corresponding 
topic map. At this point, it is enough to download the dump of Wikipedia in the very same language and make the Python code we implemented interact with such a dump.
In case a topic map features more than a language, the Python code will interface all corresponding dumps and aggregate the resulting information.\\

\section{Evaluation}
\hspace*{0.3cm} If the main goal of the algorithm described in this paper had been relatedness measurement, we could have evaluated its 
soundness and performances by relying on the typical strategy followed when an algorithm tackling relatedness measurement is designed. 
In such cases, performances are usually evaluated by computing Pearson's correlation coefficient $r$ or Spearman's correlation 
coefficient $\rho$ between the algorithm predictions and human judgements. With respect to this practice, standard datasets that are 
typically employed are, for example, the lists by Miller \& Charles~\cite{Miller91} and by Rubenstein \& Goodenough~\cite{Rubenstein65}.\\
\indent However, as already stressed in Section 1, the actual purpose of the present work is \emph{customer profiling}: not only this 
means that relatedness measurement is just a preliminary step, but it also results in the fact that, when relatedness has actually to be
measured in the process, the terms involved usually lie several levels of hypernymy far one from the other. The datasets mentioned before 
--- and similar ones --- are typically made up of terms whose hypernymy relations do not own the features we want to focus on. Consequently, in order
to apply our algorithm to such datasets, we would be compelled to completely pervert the nature of the algorithm itself and engineer
an entirely new one, drifting apart from our main goal.\\
\indent In order to evaluate the performances of our algorithm in a way consistent with its purpose and also to find the best values for 
the model parameters $\alpha$, $l_{max}$ and $l_{th}$ defined in Section 2, we decided to apply our algorithm to the topic maps extracted 
from a set $C$ made up of $N_c$ Twitter accounts\footnote{Twitter was chosen since no owner's permission is required in order to perform topic 
extraction on his/her activities on such a platform.}, for each one of which the content of the activities should be strongly oriented towards a given category 
(referred to as ``ground-truth sink" and labelled $S_{gt}$ in what follows). For example, activities on a sportsman's account are likely to 
mostly deal with sport while topics extracted from the account of an association of literary critics should be related to literature to a 
great extent. The idea is to measure how well our algorithm is capable of identifying the $S_{gt}$ ideally associated to each account when varying 
the model parameters. This measurement can be carried out by means of some indices summarizing the average degree of correctness on set $C$ as a whole.\\ 
\indent Before introducing such indices, for later convenience we label with $C_i$ the subset of $C$ made up of those accounts whose ground-truth sink is 
category $S_i$ (with $i\in\{1,2,\ldots,N_s\}$, being $N_s$ the number of categories) and with $n_i$ the cardinality, i.e., the number of elements, 
of $C_i$. After extracting a topic map from each account, setting the model parameters $\alpha,l_{max},l_{th}$ to some tentative values and 
computing the percentages in Eq.(\ref{eq1}) for each account, the indices taken into account for evaluating the quality of the predictions --- 
called \emph{score} ($SC$), \emph{rank} ($RK$) and \emph{difference} ($\Delta$)\footnote{We are currently considering the possibility of making use also of 
more standard indices like, for example, the multi-class ROC~\cite{Landgrebe07}; however, we are now assessing whether they could consistently be employed 
in the present case or not.} --- are defined\footnote{Since $SC$, $RK$ and $\Delta$ --- as well as the similar quantities in Eqs.(\ref{eq6})-(\ref{eq7}) they depend 
upon  --- are actually functions of the model parameter, a mathematically more correct way of denoting them would be $SC(\alpha,l_{max},l_{th})$, 
$RK(\alpha,l_{max},l_{th})$ and $\Delta(\alpha,l_{max},l_{th})$. Anyway, we will stick to $SC$, $RK$ and $\Delta$ to ease the notation.} as

\begin{eqnarray}
SC &=& \frac{1}{N_{s}}\sum_{i=1}^{N_{s}}sc(S_i)\ ,\nonumber\\
RK &=& \frac{1}{N_{s}}\sum_{i=1}^{N_{s}}rk(S_i)\ ,\nonumber
\end{eqnarray}
\begin{eqnarray}
\label{eq6}
\Delta &=& \frac{1}{N_{s}}\sum_{i=1}^{N_{s}}\delta(S_i)\ ,
\end{eqnarray}
\vspace*{0.3cm}

\noindent with

\vspace*{0.2cm}
\begin{eqnarray}
\label{eq7}
sc(S_i) &=& \frac{1}{n_i}\sum_{a_j\in C_i}sc(a_j)\ ,\nonumber\\
rk(S_i) &=& \frac{1}{n_i}\sum_{a_j\in C_i}rk(a_j)\ ,\nonumber\\
\delta(S_i) &=& \frac{1}{n_i}\sum_{a_j\in C_i}\delta(a_j)\ ,
\end{eqnarray}
\vspace*{0.2cm}

\noindent where $i\in\{1,2,\ldots,N_s\})$, $a_j$ labels a given Twitter account and quantities $sc(a_j)$, $rk(a_j)$ and $\delta(a_j)$ 
associated to account $a_j$ are defined as follows

\vspace*{0.4cm}

\begin{itemize}

\item $sc(a_j)$ is equal to score $pc(u_j,S_{gt}(a_j))$ defined in Eq.(\ref{eq1}), where $S_{gt}(a_j)$ is the ground-truth sink associated to account 
$a_j$ and $u_j$ is its owner;

\vspace*{0.3cm}

\item $rk(a_j)$ is the ranking of the ground-truth sink $S_{gt}(a_j)$ on the basis of the percentages computed according to Eq.(\ref{eq1}) starting from 
the topic map extracted from $a_j$: if sink $S_{gt}(a_j)$ gets the highest score, then $rk(a_j)=1$, if it obtains the second highest, then 
$rk(a_j)=2$, and so on;

\vspace*{0.3cm}

\item $\delta(a_j)$ is defined as

\vspace*{0.2cm}
\begin{equation}
\label{eq8}
\delta(a_j) = \frac{pc(u_j,S_{gt}(a_j))-pc(u_j,S_{*}(a_j))}{pc(u_j,S_{gt}(a_j))}\ ,
\end{equation} 
\vspace*{0.2cm}

where $pc(u_j,S_{*}(a_j))$ corresponds to the largest value among the percentages $pc(u_j,S_m)$'s for all $m$ categories (in case the largest value is not 
$pc(u_j,S_{gt}(a_j))$, i.e., in case the ground-truth category is not ranked first) or to the second largest value among the $pc(u_j,S_m)$'s (in case 
$S_{gt}(a_j)$ is ranked first).

\end{itemize}

\vspace*{0.3cm}

\begin{figure}  
\begin{minipage}{0.45\textwidth} 
\includegraphics[width=\linewidth]{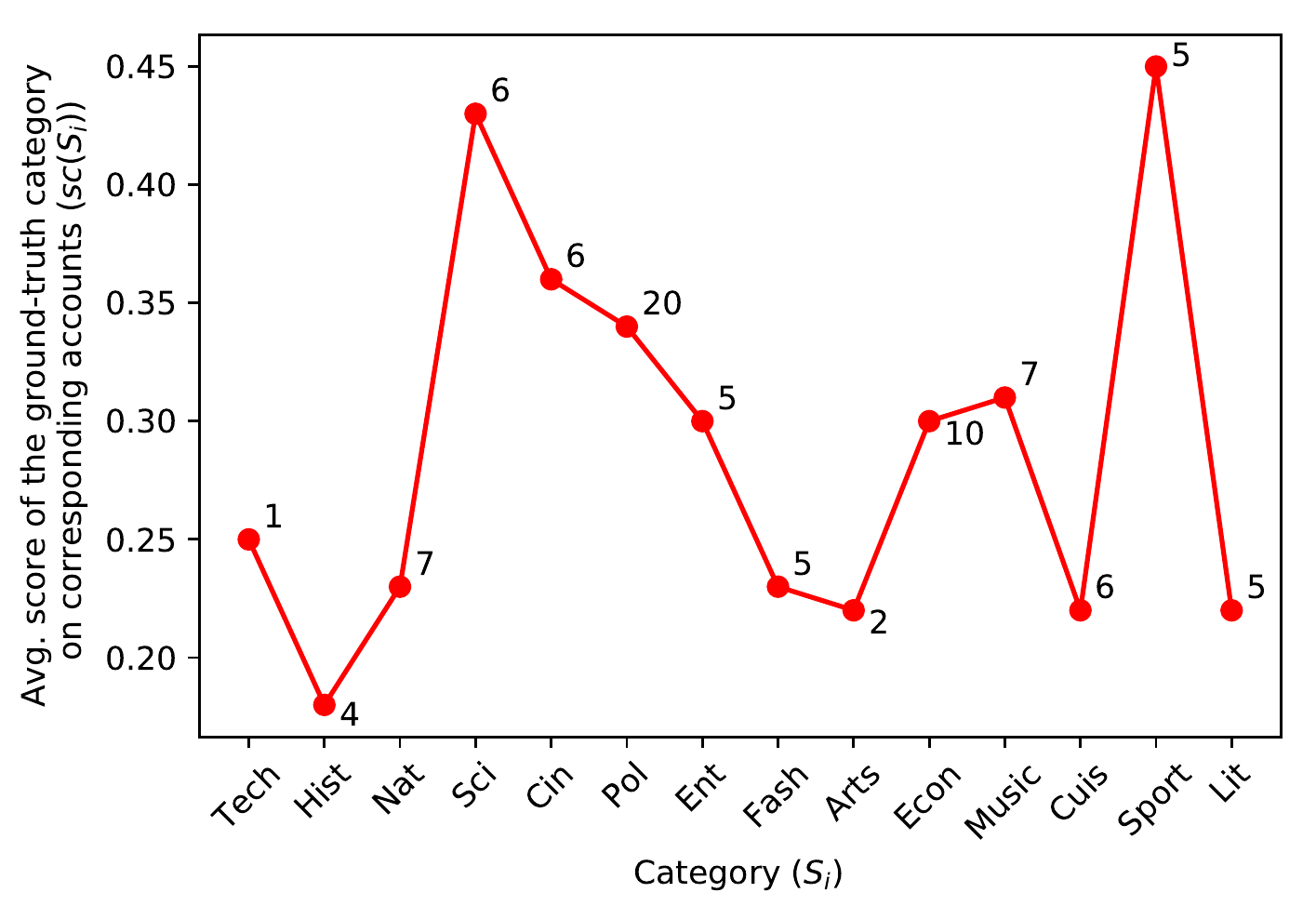}    
\end{minipage}    
\hspace{\fill}  
\begin{minipage}{0.45\textwidth} 
\includegraphics[width=\linewidth]{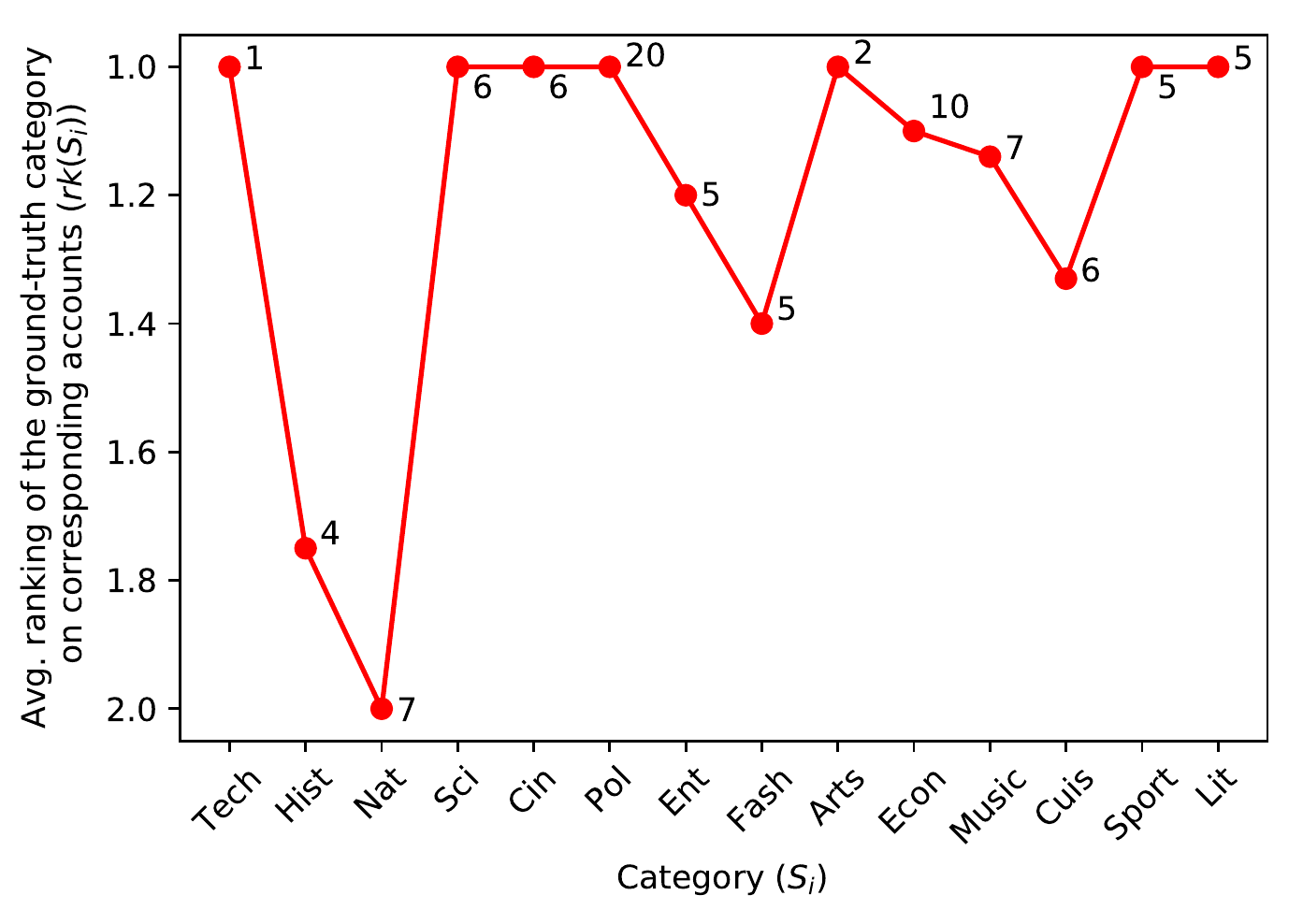}    
\end{minipage}
\begin{minipage}{0.45\textwidth} 
\includegraphics[width=\linewidth]{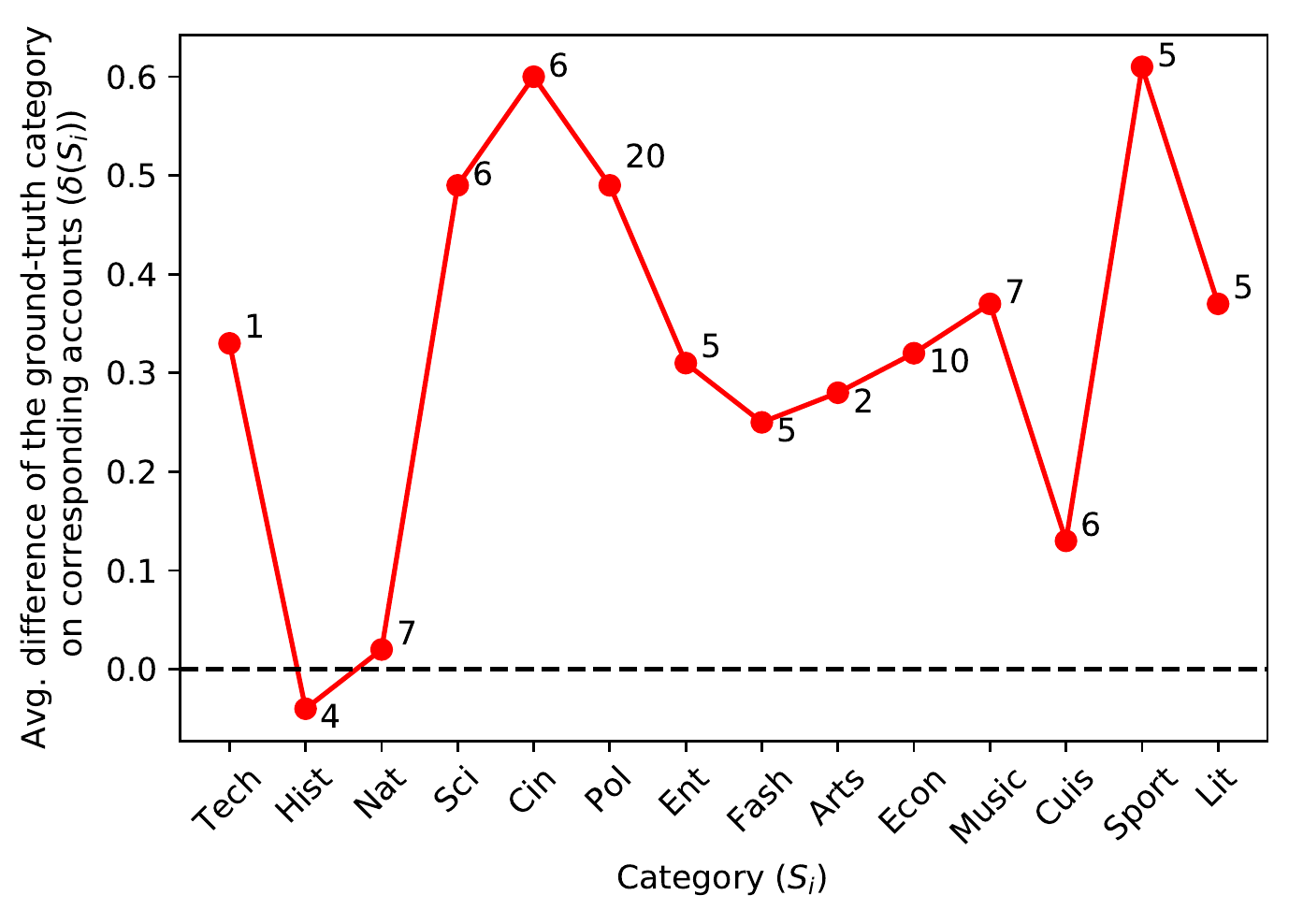}    
\caption{Plots of $sc(S_i)$, $rk(S_i)$ and $\delta(S_i)$ as defined in Eqs.(\ref{eq7}) for 14 categories. For each category $S_i$, the plots display --- from up to 
bottom --- the average score, ranking and difference of $S_i$ as computed from the topic maps extracted from the Twitter accounts that should have $S_i$ as 
ground-truth sink. The number of Twitter accounts supposedly having $S_i$ as ground-truth sink is shown close to each point.\\
\hspace*{0.3cm} The categories --- displayed on the horizontal axis of each panel --- are ``Technology and applied sciences" (Tech), ``History" (Hist), 
``Nature" (Nat), ``Science" (Sci), ``Cinema" (Cin), ``Politics" (Pol), ``Entertainment" (Ent), ``Fashion" (Fash), ``Arts" (Arts), ``Economics" (Econ), 
``Music" (Music), ``Cuisine" (Cuis), ``Sport" (Sport) and ``Literature" (Lit).} 
\label{Fig4}
\end{minipage}
\end{figure}

\indent Some remarks are now due.\\
$SC$, $RK$ and $\Delta$ are all defined through a ``double average": a first average --- that in Eqs.(\ref{eq7}) --- is computed within each 
subset $C_i$, i.e., on all Twitter accounts sharing a common ground-truth sink, while a second --- in Eqs.(\ref{eq6}) --- is evaluated on the different 
categories starting from the mean values obtained in Eqs.(\ref{eq7}). The rationale is that one wants to set all sinks on the same footing before evaluating 
score, rank and difference: if one had computed the plain average on all the $N_{c}$ topic maps, those categories that are ground truth for (relatively) many 
accounts in the set $C$ would have had more weight, thus making the indices ``artificially'' high or low, depending on how well or bad these over-represented 
categories emerge within the corresponding subsets $C_i$'s. The ``pre-average'' in Eqs.(\ref{eq7}) gives each category a single vote in Eqs.(\ref{eq6}) --- 
so to speak ---, and any ``artificial" increase (or decrease) in $SC$, $RK$ and $\Delta$ should in principle be avoided.\\
\indent As for the meaning of the indices, loosely speaking $SC$ essentially measures the average share of activities that, for each account, can be related to 
the corresponding ground-truth sink $S_{gt}$. Bearing in mind that, by virtue of Eq.(\ref{eq1}), such a share ranges from 0 to 1, in the ideal case when all 
activities are associated solely to $S_{gt}$ for all accounts, then $SC=1$; conversely, in the opposite case when no activities can be related to $S_{gt}$ on 
any account at all, $SC$ would be equal to 0. Thus, the closer $SC$ to 1, the better. However, $SC$ will never be exactly equal to 1, partly because --- in general 
--- topics are usually related to different categories (though with different degrees of relatedness), partly because it is quite natural that a
user has typically more than one interests --- there might be a sink that stands out given perhaps its relation to the user's job (we are assuming that such a 
category is $S_{gt}$), but other ones will usually be featured in his/her activities.\\
\indent In themselves, the individual $sc(a_j)$'s out of which $SC$ is eventually computed are a sort of absolute measurement, i.e. they do not convey any explicit 
information about the way $S_{gt}$ compares with the other categories that are supposed to relate less than $S_{gt}$ to the topic map extracted from a given account. 
For example, an apparently low value of $sc(a_j)$ might nevertheless correspond to the highest of the percentages $pc(u_j,S_m)$, especially for a Twitter account
whose activities are rather sparse category-wise. Conversely, the ground-truth sink might get an apparently high share --- say, slightly lower than 0.5 --- but it 
might be preceded by another category whose score is actually above 0.5. Quantities $rk(a_j)$'s entering the computation of $RK$ overcome these inconveniences since
they explicitly yield the rank of $S_{gt}$, irrespective of whether the corresponding $sc(a_j)$'s are seemingly high or low. In the ideal case when $S_{gt}$ is 
ranked first on all accounts, the overall $RK$ will be equal to 1, otherwise it will be (increasingly) higher, thus signalling that, for a (larger and larger) number
of accounts, the ground-truth sink does not actually get ranked in the first position.\\
\indent Finally, though $rk(a_j)$ provides a comparison between $S_{gt}$ and all other categories on account $a_j$, still it does not measure the gap between the 
ground-truth sink and the remaining ones. For example, on a given account $a_j$, $rk(a_j)$ will be equal to 1 if $S_{gt}$ gets ranked first, irrespective of whether 
this occurs by a large margin or by a very tiny one. On the contrary, $\delta(a_j)$ is defined in a such a way to have very different values in the case $S_{gt}$ 
clearly stands out and in the case it is ranked first by a narrow margin. In fact, according to Eq.(\ref{eq8}), $\delta(a_j)$ reads 1 when the score $pc(u_j,S_m)$ 
is zero for all categories except for the ground-truth sink (this would be the ideal, though hardly-reachable scenario), but it gets negative as soon as $S_{gt}$ is 
not ranked first --- the larger the gap between the ground-truth category and the category ranked as the first one, the more negative $\delta(a_j)$. Such features are
reflected in the overall index $\Delta$ which will read 1 in the above-mentioned ideal scenario and be (increasingly) negative as soon as, on a (wider and wider)
bunch of accounts, $S_{gt}$ is not ranked first by a large margin with respect to the leading sink.\\ 
\indent Bearing these observations in mind, we profiled $N_c = 89$ accounts altogether, shared among 14 out of the $N_s = 18$ categories\footnote{For some sinks, i.e., 
``Culture", ``Geography", ``Philosophy" and ``Religion", we did not find any suitable accounts or we were able to find only some where the amount of activities 
was too scarce to yield reliable results. These 4 sinks are obviously discarded in computing $SC$, $RK$ and $\Delta$.} we currently include in our study, trying 
to diversify --- for each category -- the kind of user. For instance, in profiling accounts related to sink ``Cinema", we considered not only accounts owned by 
actors/actresses or directors, but also those run by producers. Similarly, while focusing on category ``Music", we analyzed the accounts of artists 
belonging to different genres (rock, hip hop, etc.), of fan clubs and of those institutions --- like theaters --- often hosting music events.\\
\indent We computed $SC$, $RK$ and $\Delta$ for several setups of the model parameters $\alpha$, $l_{max}$ and $l_{th}$ and found out that the best results
are obtained with $\alpha = 3$, $l_{max} = 6$ and $l_{th} = 12$ for all the three performance indices: more precisely, we get $SC=0.29$, $RK=1.21$ and $\Delta=0.32$. 
Roughly speaking, on average $S_{gt}$ can be associated to more than one fourth of the activities of an account, it gets ranked first in 4 out of 5 cases 
(in the fifth case, it is ranked as the second sink) and its percentage is approximately $10\%$ higher than that of the strongest competitor 
sink-wise\footnote{This piece of information can be obtained by evaluating the numerator on the r.h.s. of Eq.(\ref{eq8}) after replacing the denominator with $SC$ 
and the l.h.s. with $\Delta$.}.\\
\indent For the optimal values of the model parameters quoted before, plots in Figure \ref{Fig4} show the values of observables $sc(S_i)$, $rk(S_i)$ and $\delta(S_i)$ 
defined in Eqs.(\ref{eq7}) for all 14 categories eventually taken into account in the process of performance evaluation. Each panel of Figure \ref{Fig4} displays
the categories on the horizontal axis; the upper panel shows, for each category $S$, the average score of such category as computed on the topic maps extracted 
from those Twitter accounts having $S$ as ground-truth sink. For example, on the topic maps obtained from the 6 accounts that should mostly refer to ``Science", the 
average score of ``Science" reads slightly less than 0.45. Similarly, the middle (lower) panel of Figure \ref{Fig4} displays the average ranking (difference) of each 
sink $S$ obtained from the activities of the Twitter accounts that should be oriented towards $S$. For instance, since, on the 10 accounts that should refer to 
``Economics", such a sink is ranked first on 9 accounts and second on the remaining one, $rk(``Economics")=1.1$ (i.e., $rk=(1+1+1+1+1+1+1+1+1+2)/10=1.1$), as shown 
in the middle panel of Figure \ref{Fig4}.

\begin{figure}
\centering
\includegraphics[width=\columnwidth]{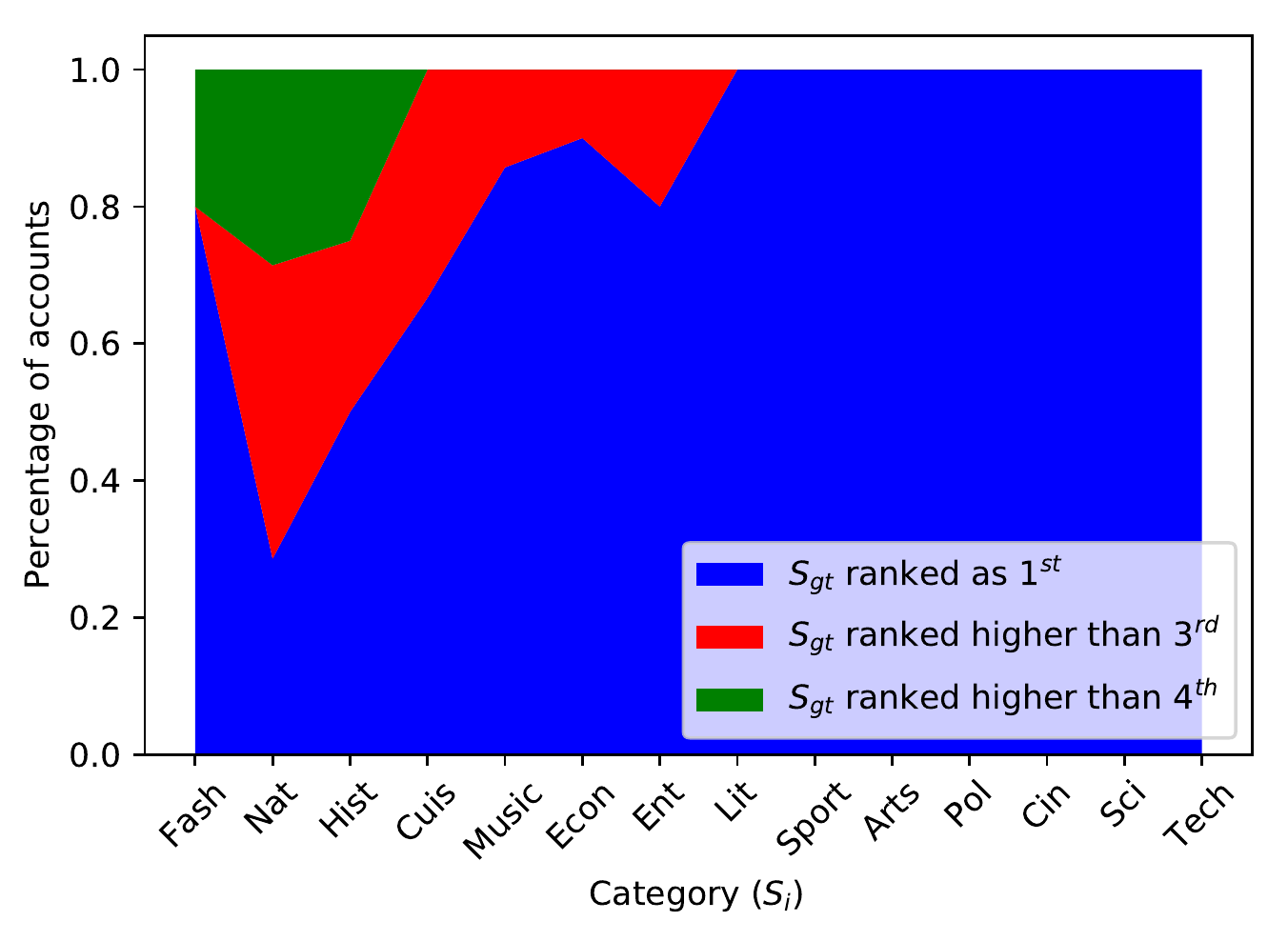}
\caption{For each category $S_i$, the percentage of accounts where $S_i$ is ranked as first (blue) or within the first two (red) or three (green) positions is 
displayed. The percentage is computed on the subset referred to in the text as $C_i$, i.e., on the subset of accounts for which $S_i$ is supposed to be 
the ground-truth category. The labels for the categories on the horizontal axis are explained in the caption of Figure \ref{Fig4}.}
\label{Fig5}
\end{figure}

\indent Figure \ref{Fig5} delves into the ranking of $S_{gt}$ within the corresponding subset $C_i$, i.e., it shows --- for each category $S_i$ --- the percentage of 
accounts, computed on the subset where $S_i$ is supposed to be the ground-truth sink, on which $S_i$ is ranked first or within the first two or three categories
on the basis of the score $pc$ defined in Eq.(\ref{eq1}). It can be seen that there are seven categories --- i.e., ``Literature", ``Sport", ``Arts", ``Politics",
``Cinema", ``Science" and ``Technology and applied sciences" --- that are ranked first on all accounts whose activities are supposedly mostly 
oriented towards them, i.e., on all accounts belonging to the correponding subset $C_i$. At worst, the ground-truth sink is ranked at the third position: this occurs
for four accounts (two of them associated to category ``Nature", one to ``History" and one to ``Fashion") while, on all other accounts, $S_{gt}$ is always ranked 
within the first two positions.\\    
\indent  For some categories, the picture looks very good. For instance, in the 5 accounts whose ground-truth sink is supposed to be ``Sport", not only this sink 
is always ranked as the first one, but, on average, 45\% of the activities can be related to it alone while the second-ranked category is associated to 
(roughly) 17\% of the extracted topics (this latter info can be loosely obtained by combining the results for ``Sport" in the upper and in the lower plot in 
Figure \ref{Fig4}). In other words, as expected ``Sport" clearly stands out in the activities extracted from the corresponding accounts. A similar situation holds 
for other sinks, like ``Science" and ``Cinema".\\
\indent On the opposite side lie categories like ``History". In fact, in the 4 accounts supposedly related to it, less than 20\% of activities can be associated
to such a sink which is ranked first only in two cases. This poor scenario is also reflected in $\delta(``History")$, which lies below the threshold
--- corresponding to the dashed line in the lower plot of Figure \ref{Fig4} --- separating the categories for which the ground-truth sink is mostly ranked first (this 
resulting in a $\delta$ with positive sign) from those for which this does not hold.\\
\indent In order to cast some light on the bad results obtained for some sinks, we are currently ``manually" checking the topic maps extracted from the accounts
supposedly related to such bad-behaving categories. By this approach, we aim at assessing to which extent the fault is in the model architecture rather than in
our initial assumptions on these accounts --- in fact, they might be (much) less related to the ground-truth sink than supposed a priori.\\

\section{Conclusions}
\hspace*{0.3cm} In this paper we described an algorithm to carry out the profiling of a user starting from his/her activities on social media,
leveraging on the so-called \emph{social login}. This algorithm makes use of the Wikipedia ontology to project the concepts
referred to in such activities onto a predefined set of categories and employs the resulting map to profile the user's interests.
A preliminary implementation --- in the form of a Python library using a dump of the Italian version of Wikipedia --- has been developed 
and its performances are currently being tested on a series of benchmark Twitter accounts whose activities should be strongly oriented 
towards a specific category.\\ 
\indent The next stage of our work will involve further checks on the model performances, the extension of the existing library in
order to be able to handle languages other than Italian and the eventual inclusion of the algorithm into a commercial software for 
customer profiling.\\   
\newline
\emph{Acknowledgements} --- This project has received funding from the European Union's Horizon 2020 research and innovation programme under grant agreement 
No 739783 (DataSci4Tapoi).
\newline
\indent The information and views set out in this study are those of the author(s) and do not necessarily reflect the official opinion of the European Union. Neither the European 
Union institutions and bodies nor any person acting on their behalf may be held responsible for the use which may be made of the information contained therein.
\newline

\bibliographystyle{ACM-Reference-Format}
\bibliography{refs}

\end{document}